\documentclass{article} 

\usepackage[utf8]{inputenc}
\usepackage[all]{xy}
\usepackage{color}
\usepackage{graphicx}

\usepackage{amsmath}
\usepackage{subfigure}
\usepackage{setspace}
\usepackage{geometry} 
\usepackage[labelsep=period]{caption}  
\usepackage[none]{hyphenat}
\usepackage{breakcites} 

\newcommand{\absolute}[1]{\left\lvert#1\right\rvert}

\newcommand{\xor}{{\rm XOR}}

\DeclareMathOperator{\dbh}{DBH}

\DeclareMathOperator{\static}{static}
\DeclareMathOperator{\scaled}{scaled}
\DeclareMathOperator{\datum}{date}
\DeclareMathOperator{\latin}{latin}
\DeclareMathOperator{\stemtag}{stemtag}
\DeclareMathOperator{\Tag}{tag}

\author{
 M. W. Jahn, P. E. Bradley
 }

\title{A scaled space-filling curve index applied to tropical rain forest tree distributions}

\begin{document}

\maketitle

\abstract{
In order to be able to process the increasing amount of spatial data, efficient methods for their handling need to be developed.
One major challenge for big spatial data is access. This can be achieved through space-filling curves, as they have the property that nearby points on the curve are also nearby in space. They are able to handle higher dimensional data, too. Higher dimensional data is widely used e.g.\ in CityGML and is becoming more and more important.
In a laboratory experiment on a  tropical rain forest tree data set of 2.5 million points taken from an $18$-dimensional space, it is demonstrated
that the recently constructed scaled Gray-Hilbert curve index performs better than its standard static version, saving a significant amount of space for a projection of the data set onto   $8$ attributes. The implementation is based on a binary tree in a data-driven process, in a similar way as e.g.\ the R-tree. Its scalability allows the handling of different kinds of data distributions which are reflected in the tree structure of the index.
The relative efficiency of the scaled Gray-Hilbert curve in comparison with the best static version is seen to depend on the distribution of the point cloud. A local sparsity measure derived from properties of the corresponding trees can distinguish point clouds with different tail distributions.
The different resulting binary trees are visualised to illustrate the influences of the different tail distributions they have been built on.
}


\section{Introduction}

Big spatial data requires efficient methods for their processing.
Not only is the amount of geographical data increasing, but also their dimensionality, as 
other geographical attributes found in e.g.\ forest cadastres, 3D City models or infrastructure models etc.\ become relevant. 
As the number of points 
becomes large and lives inside a multi-dimensional space, efficient access is becoming a major issue
 for big spatial data. In order to handle such issues,
space-filling curves have turned out to be a suitable indexing method, as nearby points on the curve are 
nearby in the space. Hilbert curves are a special case with particularly good self-similarity properties which made them popular in the domain of computer science. With geo-spatial data, they also have been successfully used.
However,  only static space-filling curves have been used so far, and these reportedly cause storage problems as dimension increases. In \cite{p-adicHilbert}, a scaled index based on a $p$-adic generalisation of higher-dimensional Hilbert curves was developed. The number of iterations there differs locally and depends on the local density of the point cloud. This index is applied in the present article to rain forest data consisting of 2.5 million data points in a multi-dimensional space, and is also compared with the traditional static Hilbert curve index.

\smallskip
The contributions of this article are
\begin{itemize}
    \item the implementation of a data-driven index structure based on the index inherent in space-filling curves themselves
    \item to exploit the inherent tree-structure of space-filling curves for $nD$ point clouds
    \item an index which can handle different kinds of distributions through its scalability
    \item reduced storage by not storing space-filling curve keys
    \item a study of the influence of point cloud distributions on the tree structure of the index 
\end{itemize}

\smallskip
The following section gives an overview of the state of the art of space-filling curve indices. That section is followed by a presentation of our methodology, in which the scaled Hilbert curve index is explained. Section 4 contains a description of the rain forest tree data set. The article concludes with a section presenting the experimental results.

\section{State of the art}

Motivated by Cantor's counter-intuitive result which says that the unit interval and Euclidean $n$-dimensional space have the same cardinality,
Peano proved in the year 1890 that there exists a continuous mapping of the unit interval onto the unit square.  This means the discovery of a curve passing through all points of the unit square \cite{Peano1890}. 
A year later, Hilbert published a variation of Peano's construction by using a binary subdivision in order to approximate the curve \cite{Hilbert1891}. This \emph{Hilbert curve} is now widely used in computer science. 

\smallskip
The above are examples of so-called \emph{space-filling curves} which can be generalised to higher dimension. Analogues 
 of Hilbert curves for higher dimension use the binary reflected Gray code which is a special re-ordering of the binary numbers
\cite{HR2007,Haverkort2017}. In \cite{p-adicHilbert}, a systematic approach to $n$-dimensional Hilbert curves and to their $p$-adic generalisations is undertaken, and thus a scaled index for 
 data taken from a space of high dimension is obtained, and tested for dimension up to 431. 

\smallskip
In \cite{MAK2002}, the performance of various different space-filling curves with respect to different criteria is studied. Due to the continuity of the space-filling curve, nearby points on the curve correspond to nearby points in the space indexed by the curve. This makes space-filling curves interesting for spatial information.
Applications in this domain include city models in 3D, where it has been observed that they allow for a more efficient information retrieval  than the  CityGML format \cite{UASRM2013}. 
The management of space-time point clouds using space-filling curves has been studied in \cite{POTB2016}, where a B-tree index is imposed onto space-filling curve keys of space-time data points, which is a 2-step approach. A similar two-step approach using an $n$-dimensional space-filling curve library was proposed and tested for point clouds in  4D using the Oracle Index Organized Table instead of the B-tree \cite{GOC2018}.
In geo-informatics, already a Hilbert curve index for data in 6D space has been observed as computationally challenging \cite{VCLB2018}.

\section{Methodology}

The \emph{binary reflected Gray code} is an ordering of the binary numbers in such a way that subsequent numbers differ only by one bit \cite{Gray1953}.
According to \cite{Knuth2004}, it can be represented by the map
\[
x\mapsto x+(x\triangleright 1)
\]
on binary numbers, where $x\triangleright k$ means the right-shift of $x$ by $k$ places, and $+$ is addition modulo $2$ or, equivalently, the logical $\xor$-operator. 

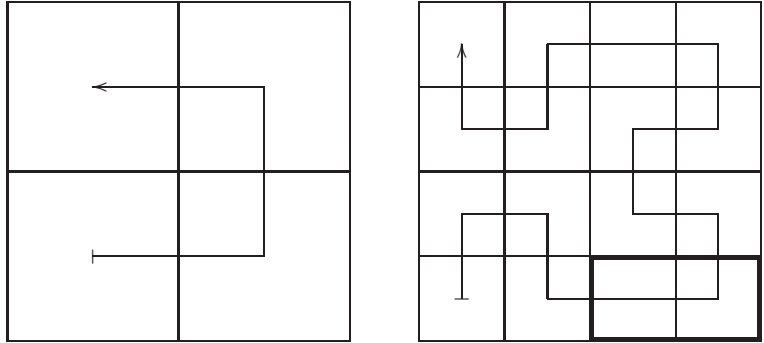
\begin{figure*}
\[
\xymatrix@=26pt{
*\txt{}\ar@{-}[rrrr]\ar@{-}[dddd]&&*\txt{}\ar@{-}[dddd]&&
*\txt{}\ar@{-}[dddd]\\
&*\txt{}&&*\txt{}\ar@{->}[ll]&\\
*\txt{}\ar@{-}[rrrr]&&&&*\txt{}\\
&*\txt{}\ar@{|-}[rr]&&*\txt{}\ar@{-}[uu]&\\
*\txt{}\ar@{-}[rrrr]&&*\txt{}&&*\txt{}
}
\qquad
\xymatrix@=10pt{
*\txt{}\ar@{-}[rrrrrrrr]\ar@{-}[dddddddd]&&*\txt{}\ar@{-}[dddddddd]&&*\txt{}\ar@{-}[dddddddd]&&*\txt{}\ar@{-}[dddddddd]&&*\txt{}\ar@{-}[dddddddd]\\
&*\txt{}&&*\txt{}\ar@{-}[dd]&&*\txt{}\ar@{-}[ll]&&*\txt{}\ar@{-}[ll]&\\
*\txt{}\ar@{-}[rrrrrrrr]&&&&&&&&*\txt{}\\
&*\txt{}\ar@{->}[uu]&&*\txt{}\ar@{-}[ll]&&*\txt{}\ar@{-}[rr]&&*\txt{}\ar@{-}[uu]&\\
*\txt{}\ar@{-}[rrrrrrrr]&&&&*\txt{}&&*\txt{}&&*\txt{}\\
&*\txt{}\ar@{-}[rr]&&*\txt{}\ar@{-}[dd]&&*\txt{}\ar@{-}[uu]&&*\txt{}\ar@{-}[ll]&\\
*\txt{}\ar@{-}[rrrrrrrr]&&&&*\txt{}\ar@[blue]@{-}[dd]\ar@[blue]@{-}@<.5pt>[dd]\ar@[blue]@{-}@<1pt>[dd]&&&&*\txt{}\ar@[blue]@{-}[llll]\ar@[blue]@{-}@<.5pt>[llll]\ar@[blue]@{-}@<1pt>[llll]\\
&*\txt{}\ar@{|-}[uu]&&*\txt{}\ar@{-}[rr]&&*\txt{}\ar@{-}[rr]&&*\txt{}\ar@{-}[uu]&\\
*\txt{}\ar@{-}[rrrrrrrr]&&*\txt{}&&*\txt{}\ar@[blue]@{-}[rrrr]\ar@[blue]@{-}@<.5pt>[rrrr]\ar@[blue]@{-}@<1pt>[rrrr]&&*\txt{}&&*\txt{}\ar@[blue]@{-}[uu]\ar@[blue]@{-}@<.5pt>[uu]\ar@[blue]@{-}@<1pt>[uu]
}
\]
\caption{The two-dimensional binary reflected Gray code (left),
and the second iteration of the two-dimensional Gray-Hilbert curve (right).}\label{Graycode2dim}
\end{figure*}

\smallskip
If $I^n$ is the $n$-dimensional hypercube, then an $n$-digit binary number represents a subhypercube of $I^n$ obtained by 
applying a subdivision into two equal parts to each coordinate, and each $0$ means the first, and $1$
the second half. The binary reflected Gray code then yields a traversal of all $2^n$ subhypercubes of $I^n$ by moving along neighbouring subhypercubes in such a way that the binary codes of the first and the last ones differ in precisely one coordinate. For $n=2$, this is depicted in Figure \ref{Graycode2dim} (left).

Now, if $I^n$ is iteratively further subdivided, then there exist for each iteration $k$ transformations of the binary reflected Gray code for each subhypercube such that the union of all these transformed traversals form a curve which passes through all the $2^{kn}$ subhypercubes, i.e.\ the binary codes of any subsequent sub\-hyper\-cubes   differ in precisely one coordinate. Such a curve is a so-called \emph{Hilbert curve}, or \emph{Gray-Hilbert curve}, as it is called in \cite{p-adicHilbert}. 

\smallskip
In the  article above, only affine transformations are allowed. It turns out that these are given by a permutation of coordinates followed by a translation, so that there are $(n-1)!$ possible transformed Gray codes after fixing the start point and the  end point, which are neighbours \cite{p-adicHilbert}. 
As each start and end point is determined by the previous iteration, 
we see from this that the $2$-dimensional Gray-Hilbert curve is uniquely determined by the Gray code and coincides with the well-known Hilbert curve whose second iteration is depicted in Figure \ref{Graycode2dim} (right).

\begin{figure*}
\[
\xymatrix@=8pt{
&&&&&&&&&&&&&&&&
*\txt{$\bullet$}\ar@[blue]@{-}[dllllllll]\ar@[blue]@<.5pt>@{-}[dllllllll]
\ar@{-}[drrrrrrrr]
\\
&&&y\ar@{--}[rrrrrrrrrrrrrrrrrrrrrrrrr]&&&&&*\txt{$\bullet$}
\ar@{-}[dllll]\ar@[blue]@{-}[drrrr]\ar@[blue]@{-}@<.5pt>[drrrr]&&&&&&&&&&&&&&&&*\txt{$\bullet$}\ar@{-}[dllll]\ar@{-}[drrrr]&&&&
\\
&&&x\ar@{--}[rrrrrrrrrrrrrrrrrrrrrrrrr]&*\txt{$\bullet$}\ar@{-}[dll]\ar@{-}[drr]&&&&&&&&
*\txt{$\bullet$}\ar@[blue]@{-}[dll]\ar@[blue]@{-}@<.5pt>[dll]\ar@{-}[drr]&&&&&&&&*\txt{$\bullet$}\ar@{-}[dll]\ar@{-}[drr]&&&&&&&&*\txt{$\bullet$}\ar@{-}[dll]\ar@{-}[drr]
\\
x\ar@{--}[rrrrrrr]&&*\txt{$\bullet$}\ar@{-}[dl]\ar@{-}[dr]&&&&*\txt{$\bullet$}\ar@{-}[dl]\ar@{-}[dr]&&y\ar@{--}[rrrrrrr]&&*\txt{$\bullet$}\ar@{-}[dl]\ar@{-}[dr]&&&&*\txt{$\bullet$}\ar@{-}[dl]\ar@{-}[dr]&&y\ar@{--}[rrrrrrr]&&*\txt{$\bullet$}\ar@{-}[dl]\ar@{-}[dr]&&&&*\txt{$\bullet$}\ar@{-}[dl]\ar@{-}[dr]&&x\ar@{--}[rrrrrrr]&&*\txt{$\bullet$}\ar@{-}[dl]\ar@{-}[dr]&&&&*\txt{$\bullet$}\ar@{-}[dl]\ar@{-}[dr]&
\\
y\ar@{--}[rrrrrrr]&*\txt{$\bullet$}&&*\txt{$\bullet$}&&*\txt{$\bullet$}&&*\txt{$\bullet$}&x\ar@{--}[rrrrrrr]&*\txt{$\bullet$}&&*\txt{$\bullet$}&&*\txt{$\bullet$}&&*\txt{$\bullet$}&x\ar@{--}[rrrrrrr]&*\txt{$\bullet$}&&*\txt{$\bullet$}&&*\txt{$\bullet$}&&*\txt{$\bullet$}&y\ar@{--}[rrrrrrr]&*\txt{$\bullet$}&&*\txt{$\bullet$}&&*\txt{$\bullet$}&&*\txt{$\bullet$}
}
\]
\caption{The first two iterations of the Gray-Hilbert tree in dimension two with local level labellings and a specified path corresponding to the blue rectangle in Figure \ref{Graycode2dim} (right).}\label{GHT2}
\end{figure*}
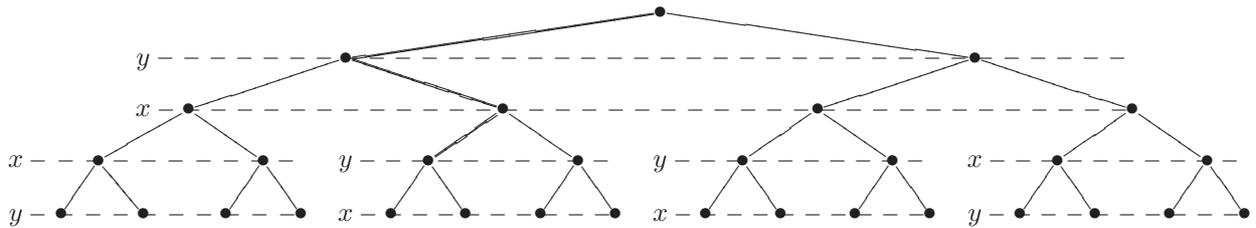

A point in the $k$-th iteration of a Gray-Hilbert curve can be represented by a sequence of $n$-digit binary numbers 
\[
(a_1,\dots,a_k)
\]
of length $k$,
where each $a_i$ signifies the (transformed) Gray code value corresponding to a hypercube at iteration $i$. The binary number $a_i$ itself corresponds to an assignment of values $0$ or $1$ to the coordinates in some specified ordering, depending on the transformation. Thus we can represent the hypercubes at level $k$ by a path from root to a leaf in a binary tree with $2^{nk}$ leaves. This tree is obtained by iteratively gluing the roots of copies of a binary tree with $n$ levels to the leaves of a tree starting with a root  vertex. 
The gluing corresponds to the hypercube subdivision.
The levels of each of the tree pieces are each labelled by a coordinate which is given by the local ordering of coordinates which is determined by the transformation used for building up the Gray-Hilbert curve at this particular hypercube subdivision. By taking an infinite iteration process, we call the resulting locally finite rooted tree the
\emph{Gray-Hilbert tree in dimension $n$}.
Figure \ref{GHT2} shows a part of the Gray-Hilbert tree of dimension two, namely the tree obtained by performing the first two iterations. 

\smallskip
A path in the Gray-Hilbert tree from root to a given vertex corresponds to a sequence of subdivisions of certain coordinates which yields a subcuboid of $I^n$ whose edges are one of two kinds: either short or long.  The blue path in Figure \ref{GHT2} thus corresponds to the blue rectangle in Figure \ref{Graycode2dim} (right).

\smallskip
Let $S\subset I^n$ be a finite point cloud. The \emph{Gray-Hilbert tree generated by $S$} is defined as being the smallest subtree of the Gray-Hilbert tree in dimension $n$ having the same root, and also whose leaves correspond to subcuboids of $I^n$ each containing precisely one element of $S$. In \cite{p-adicHilbert}, an efficient algorithm is given for computing the Gray-Hilbert tree generated by any finite point cloud in $I^n$, which uses certain types of coordinate permutations.   
The linear ordering imposed on $S$ by identifying $S$ with the leaves of the Gray-Hilbert tree generated by $S$ is called the  \emph{scaled Gray-Hilbert curve index} for $S$.

\smallskip
In contrast to that, the traditional \emph{static Gray-Hilbert curve index} is the linear ordering of a finite point cloud $S$ given by an iteration of a Gray-Hilbert curve which uses a subdivision of $I^n$ which separates all points of $S$. 

\smallskip
In this article, we consider the two different Gray-Hilbert curve indices determined by the following permutations. The first one, called \emph{bubble} is defined in the Gray-Hilbert tree at the corresponding local root vertex of the following $n$-level binary tree
by the permutation 
\[
\begin{pmatrix}
n-1&n-2&\dots &d&d-1&\dots&1&0\\
d&n-1&\dots&d+1&d-1&\dots&1&0
\end{pmatrix}
\]
where $d$ is the coordinate number coming in the first level after that particular local root vertex.
The second type is called \emph{ring} and is given by the permutation
\[
\begin{pmatrix}
n-1&n-2&\dots &n-(d+1)&\dots&0\\
d&d-1&\dots&0&\dots&d+1
\end{pmatrix}
\]
at each entry point into a new copy of the $n$-level binary tree.

\section{The data set}

The data set represents tree-specific information taken from ca.\ 938,000 trees in Barro Colorado Island (Panama) in a series of 7 censuses, beginning in 1982, in an area of 50 hectares \cite{HCF2010,Condit1998,HFBHCWWL1999}. 
There are $18$ attributes, from which are chosen for the indexing methods $8$ attributes which seem interesting from a geo-information point of view. These are namely
the following: 
\begin{align*}
&\text{$x$-coordinate $x$ (60,035 values)} 
\\&\text{$y$-coordinate $y$ (47,099 values)} 
\\&\text{date of measurement $\datum$ (2,328 values)}
\\&\text{diameter at breast height $\dbh$ (1,823 values)} 
\\&\text{height-of-measure $\hom$ (286 values)}
\\&\text{latin name of species $\latin$ (325 values)}
\\&\text{tag number of the individual stem $\stemtag$ (152 values)}
\\&\text{tag number used in the field $\Tag$ (423,617 values)}
\end{align*}
The non-numerical attribute $\latin$ is coded by assigning  to each species a random value in the interval $[0,1]$.

\smallskip
Thus the data set used here consists of a point cloud 
of 
2.45 million data points in an $8$-dimensional space, 368,000 of which have missing values. Whereas in \cite{TJDWH2015}, sphere packings were used, it is the aim of this article to compare the performance of space-filling curve indices with respect to 
the influence of certain coordinates on the resulting tree structures.

\section{Results}

The implementation of the index methods is based on a tree structure
and is now part of the geo-spatio-temporal database architecture DB4GeO \cite{DB4Geo} which has been redesigned 
in order to 
be able to use OLTP and OLAP for
 big data issues, 
using property graph database methods \cite{JBMB2017}. The corresponding algorithm is presented in \cite{p-adicHilbert}. Tree structures are very well suited for modelling the scaled Gray-Hilbert curve and can also be applied to the static version.
The length of the path from root to leaf node equals the number of bits used to address the sub-hypercuboid which corresponds to the leaf node. 
Traversing the tree in pre-order defines a linear ordering of leaf nodes which corresponds to the linear ordering of the hypercuboids given by the Gray-Hilbert curve.

\smallskip
Given some multi-dimensional point cloud,
the leaf nodes of the Gray-Hilbert tree, or any sub-tree of it, are viewed as \emph{buckets} which are respectively \emph{empty}, \emph{filled}, \emph{underfilled}, or \emph{overfilled}, 
 depending on if the corresponding hypercuboid contains no, a specified number, less than, or more than a pre-specified number of data points. 
This specified parameter is denoted here as $s$, and is called the \emph{bucket capacity}.

\smallskip
Following \cite{p-adicHilbert},
we define the quantities 
\begin{align*}
\Omega(T,S)&=(1+\omega(T,S))
\times\text{$\#$ leaf nodes of $T$}
\\
\omega(T,S)&=
\frac{\text{$\#$ overfilled leaf nodes of $T$}}
{\text{$\#$ non-empty leaf nodes of $T$}}
\end{align*}
where $T$ is a sub-tree of the Gray-Hilbert tree and $S$ is a point cloud.

\smallskip
The quantity $\Omega(T,S)$ is a measure which takes into account the capacity of $T$, which is the number of leaf nodes, and a factor 
$1+\omega(T,S)$ where $\omega(T,S)$ is the number of leaf node splittings which would occur for $S$, if the depth of the tree $T$ were not bounded.
We call $\Omega(T,S)$ the \emph{capacity of $T$ for $S$}.

\smallskip
We will write $T_{\scaled}$ for the Gray-Hilbert tree generated by $S$.
The tree
\[
T_{\static}
\]
is defined as the finite Gray-Hilbert tree with maximal iteration number
\[
k=\left\lceil\frac{\log_2\dfrac{\absolute{S}}{s}}{n}\right\rceil
\]
as in \cite{p-adicHilbert}. Here, $\absolute{S}$ means the number of elements of $S$, and $s$ is the bucket capacity. Choosing this value of $k$ yields an optimal static Gray-Hilbert curve index with the property that, on average, no leaf node is overfilled. 

\begin{figure*}
\centering
\includegraphics[scale=.39]{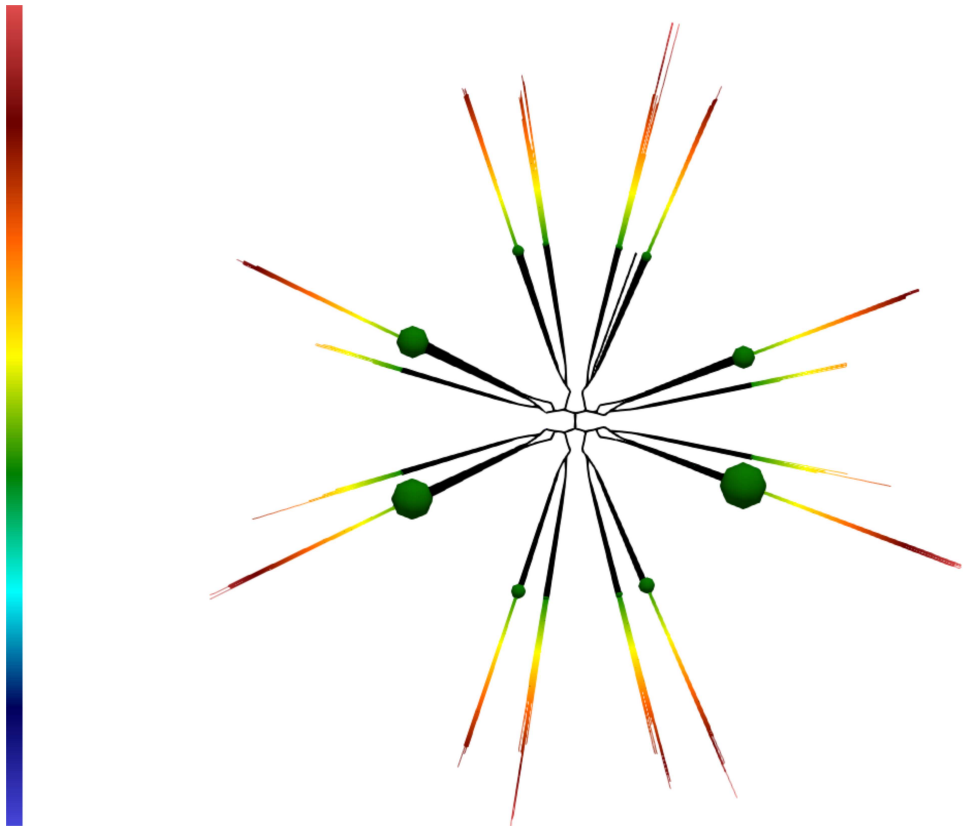}
\caption{The Gray-Hilbert trees for the projection onto the eight attributes (bubble); $T_{\static}$ in black, $T_{\scaled}$ coloured. The colours represent the levels of nodes.}
\label{GHT_BCI}
\end{figure*}


\begin{figure*}
\centering
\includegraphics[scale=.412]{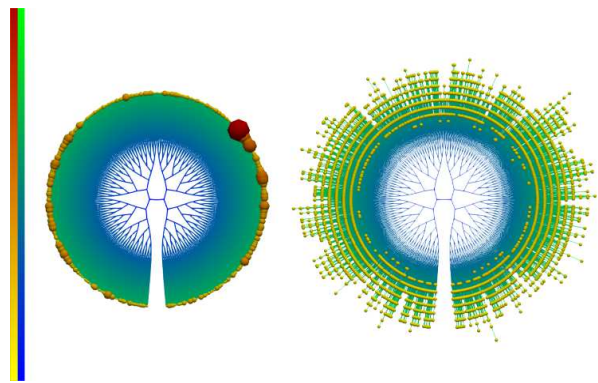}
\caption{The Gray-Hilbert trees for the projection onto $x,y,\latin$ (bubble);
$T_{\static}$ is shown on the left, and $T_{\scaled}$ on the right. Colours blue to green represent the levels, and colours yellow to red as well as the size of the glyphs represent the number of objects inside a leaf node.}
\label{GHT_BCI_xylatin}
\end{figure*}

\smallskip
Figures \ref{GHT_BCI} 
and \ref{GHT_BCI_xylatin} show the Gray-Hilbert tree for the projection onto the eight chosen attributes, 
and onto $x,y\latin$, respectively. In Figure \ref{GHT_BCI}, $T_{\static}$ is shown in black, whereas the tree $T_{\scaled}$ is coloured.
The size of the green glyphs represents the number of objects inside a leaf node of $T_{\static}$. 
In Figure \ref{GHT_BCI_xylatin},
$T_{\static}$ is shown on the left, and $T_{\scaled}$ on the right. Colours blue to green represent the levels, and colours yellow to red as well as the size of the glyphs represent the number of objects inside a leaf node.
The levels are represented by colours.

\smallskip
The iteration numbers of $T_{\static}$ are $k=3$ for all 8 attributes, 
and $k=5$ for $x,y,\latin$.

\smallskip
In order to compare the scaled and the static Gray-Hilbert curve indices, we consider the \emph{Gray-Hilbert capacity ratio for $S$}
\[
R(S)=\frac{\Omega(T_{\scaled},S)}{\Omega(T_{\static},S)}
\]
defined in \cite{p-adicHilbert}.
If $R(S)<1$, then the scaled index is more storage efficient than the optimal static index, otherwise not.

\smallskip
In \cite{p-adicHilbert}, it was mathematically proven that $R(S)$ 
is indeed never greater than $1$ if $n$ is sufficiently large, and also  tends to zero for increasing dimension $n$. In Table \ref{capacity_ratio}, it is confirmed that for the projection of the data onto all eight chosen attributes, the Gray-Hilbert capacity ratio is indeed less than one, also for quite large values of the bucket capacity $s$ (although for $s=2^{14}$ it does get close to $1$). 

\begin{table}
\centering
\begin{tabular}{|c|cc|}\hline
$\log_2 s$&\emph{bubble}&\emph{ring}\rule{0pt}{3.5mm}\\\hline
0&0.076&0.076\rule{0pt}{3.5mm}\\
1&0.052&0.052\\
2&0.032&0.032\\
3&0.018&0.018\\
4&0.010&0.010\\
5&0.006&0.005\\
6&0.644&0.649\\
7&0.342&0.355\\
8&0.180&0.185\\
9&0.097&0.092\\
10&0.051&0.050\\
11&0.026&0.026\\
12&0.015&0.014\\
13&0.009&0.007\\
14&0.912&0.804\\\hline
\end{tabular}
\caption{Values of of the Gray-Hilbert capacity ratio for the projection onto all eight variables.}\label{capacity_ratio}
\end{table}

\smallskip
The \emph{Gray-Hilbert local sparsity measure} $\rho(S,s)$ is defined via the  relationship
\[
(2s)^{\rho(S,s)}=\frac{\absolute{S}}
{\absolute{L(T_{\scaled})}}
\left(1+\omega(T_{\static})\right)
\]
where $\absolute{L(T_{\scaled})}$ is the number of leaf nodes of $T_{\scaled}$.
This  measure $\rho(S,s)$ provably takes values between $0$ and $1$ \cite{p-adicHilbert}.
If $\rho(S,s)$ is near zero, then $S$ is more likely to be uniformly distributed, whereas $\rho(S,s)$ near one indicates that this should not be the case.

\begin{table*}
\centering
\begin{tabular}{|c|cc|cc|cc|}\hline
&\multicolumn{2}{c|}{$s=1$}&\multicolumn{2}{c|}{$s=2$}&\multicolumn{2}{c|}{$s=4$}\rule{0pt}{3.5mm}\\
attributes&\emph{bubble}&\emph{ring}\rule{0pt}{3.5mm}&\emph{bubble}&\emph{ring}&\emph{bubble}&\emph{ring}\\\hline
all eight&0.928&0.928&0.744&0.744&0.729&0.729\rule{0pt}{3.5mm}\\\hline
$x,y$&{\bf 0.288}&{\bf 0.288}&{\bf 0.519}&{\bf 0.519}&{\bf 0.521}&{\bf 0.521}\rule{0pt}{3.5mm}\\\hline
$x,y,\datum$&0.886&0.886&0.890&0.776&0.790&0.790\rule{0pt}{3.5mm}\\
$x,y,\dbh$&0.858&0.858&0.691&0.691&0.680&0.680\\
$x,y,\hom$&0.857&0.857&0.745&0.745&0.702&0.702\\
$x,y,\latin$&{\bf 0.437}&{\bf 0.437}&{\bf 0.547}&{\bf 0.547}&{\bf 0.573}&{\bf 0.573}\\
$x,y,\stemtag$&0.732&0.732&0.738&0.739&0.734&0.736\\
$x,y,\Tag$&{\bf 0.693}&{\bf 0.693}&{\bf 0.698}&{\bf 0.698}&{\bf 0.678}&{\bf 0.678}\\\hline
\end{tabular}
\caption{Values of $\rho$ for projections of the data onto certain sets of attributes, and different bucket capacitys $s$. The values in boldface have almost identical tail distributions (cf.\ Figure \ref{taild}).}
\label{rho}
\end{table*}

\begin{figure*}
\centering
\begin{tabular}{cc}
\includegraphics[scale=.4]{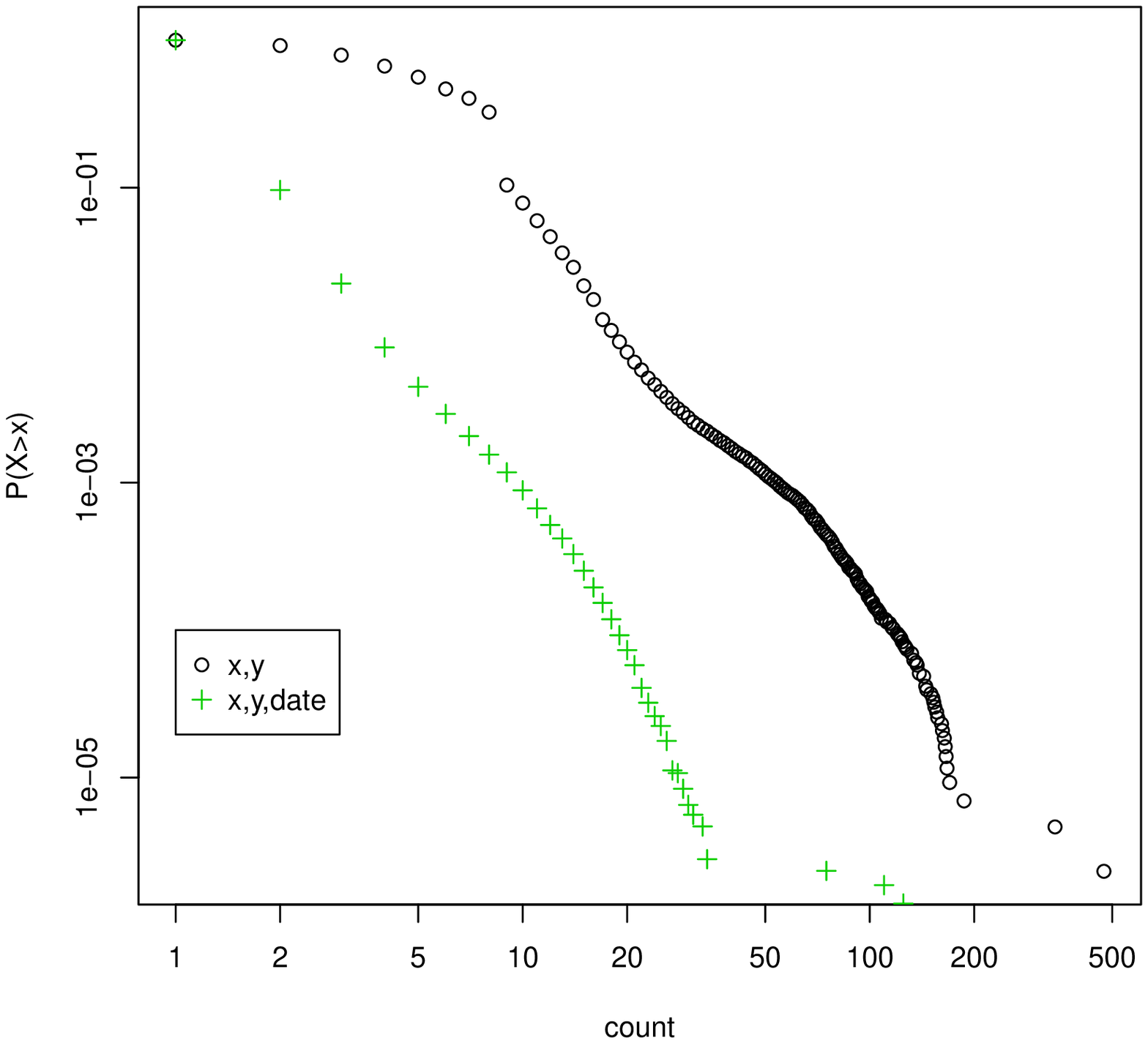}&\includegraphics[scale=.4]{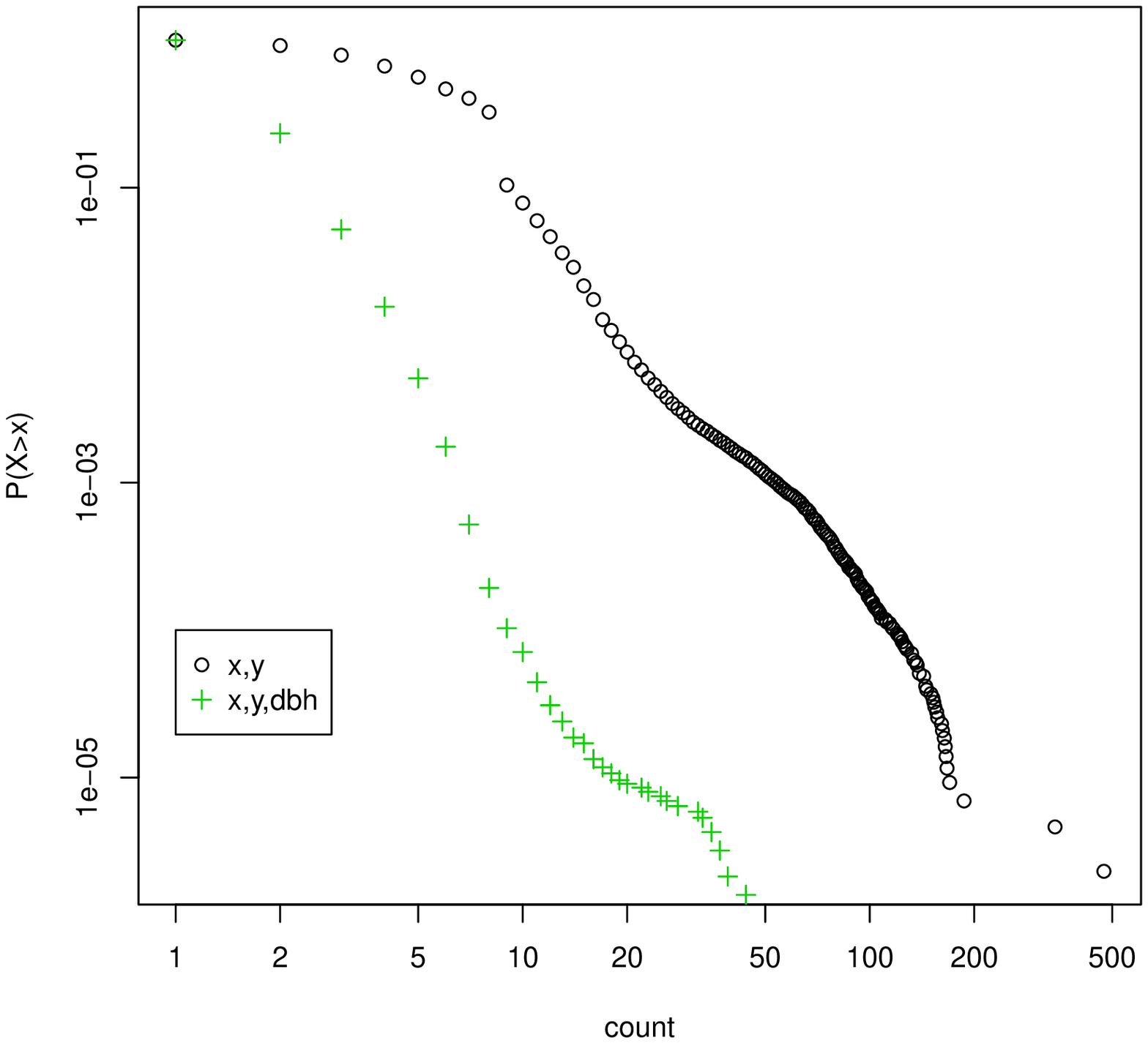}
\\
\includegraphics[scale=.4]{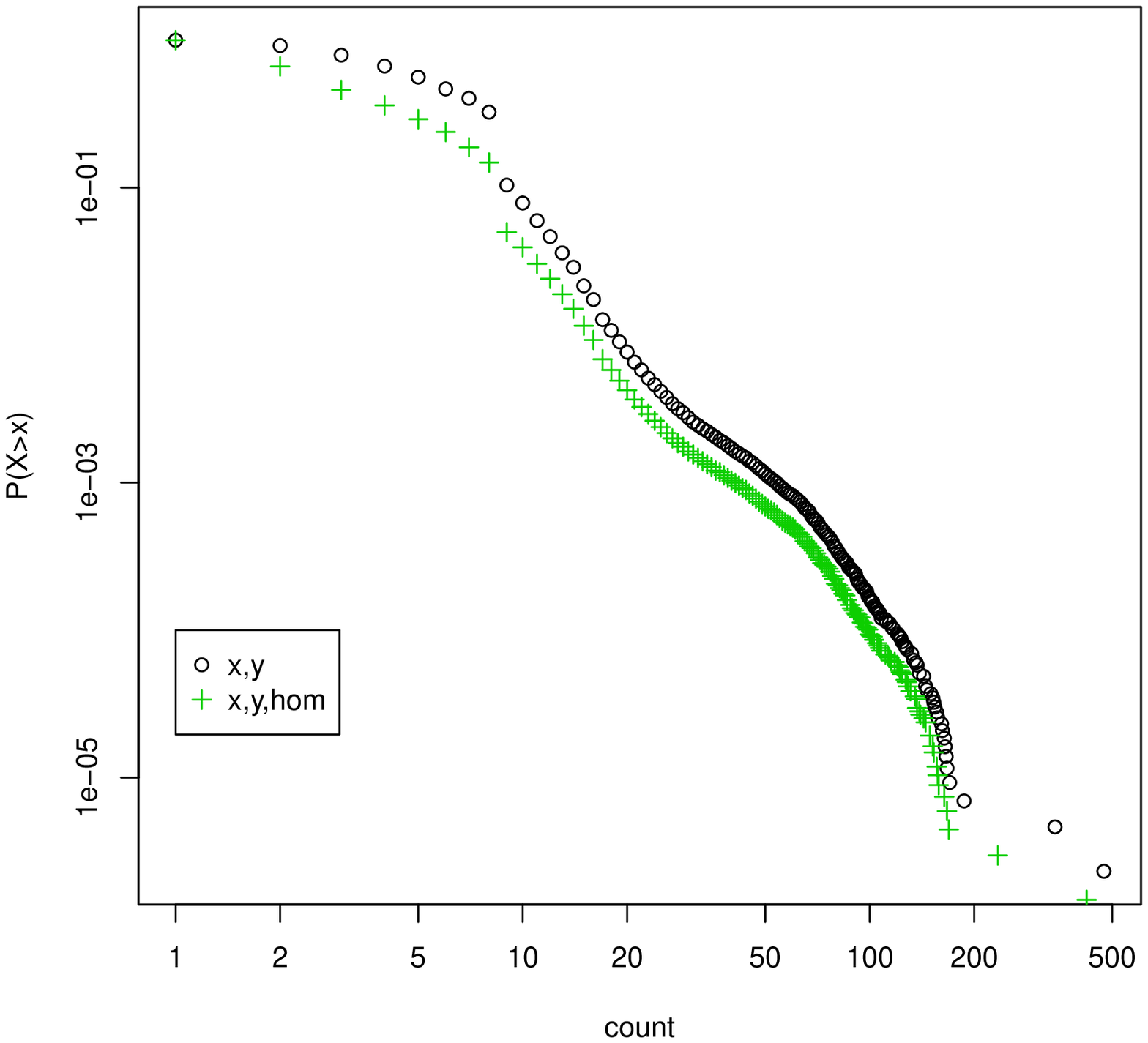}&\includegraphics[scale=.4]{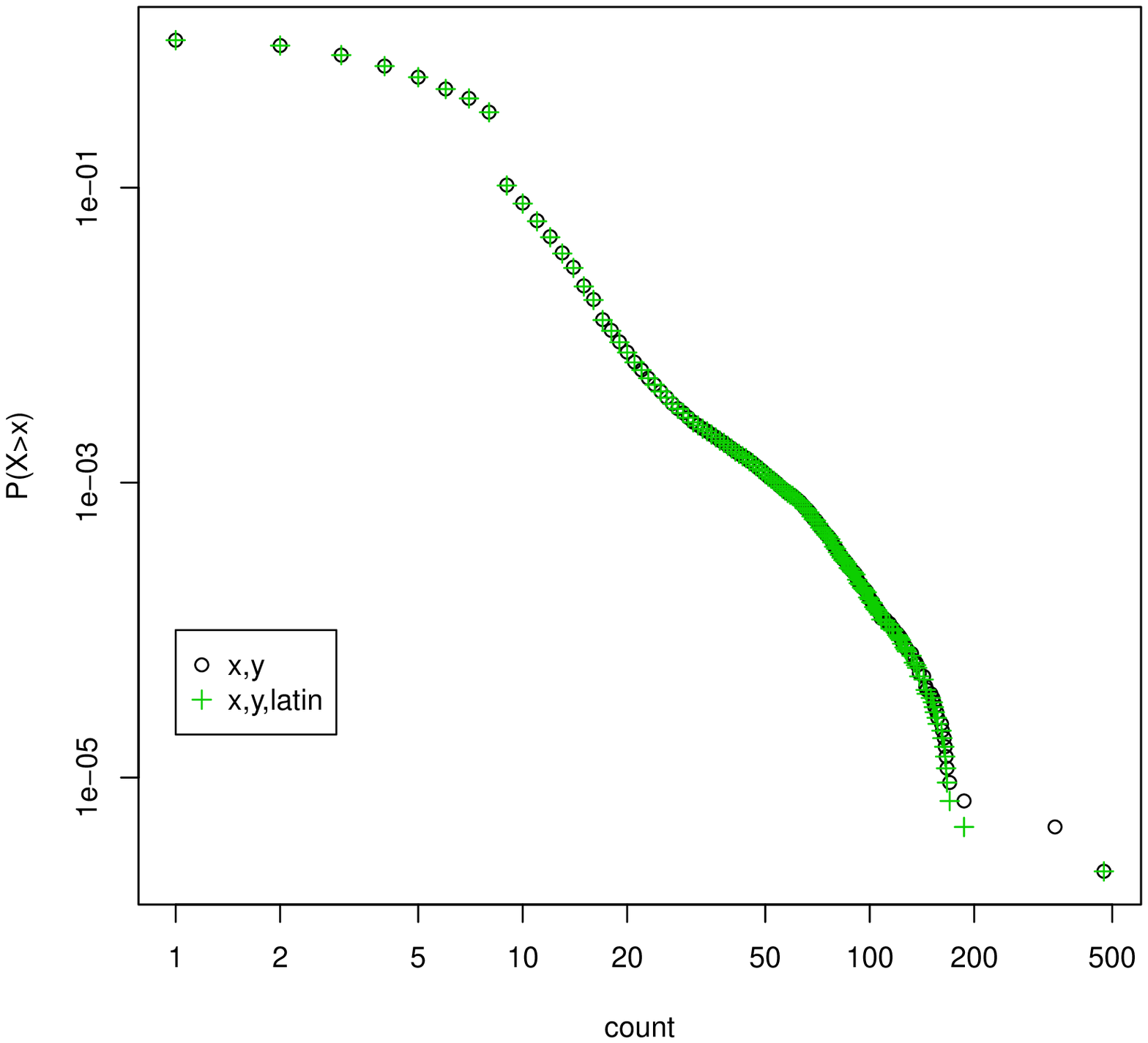}
\\
\includegraphics[scale=.4]{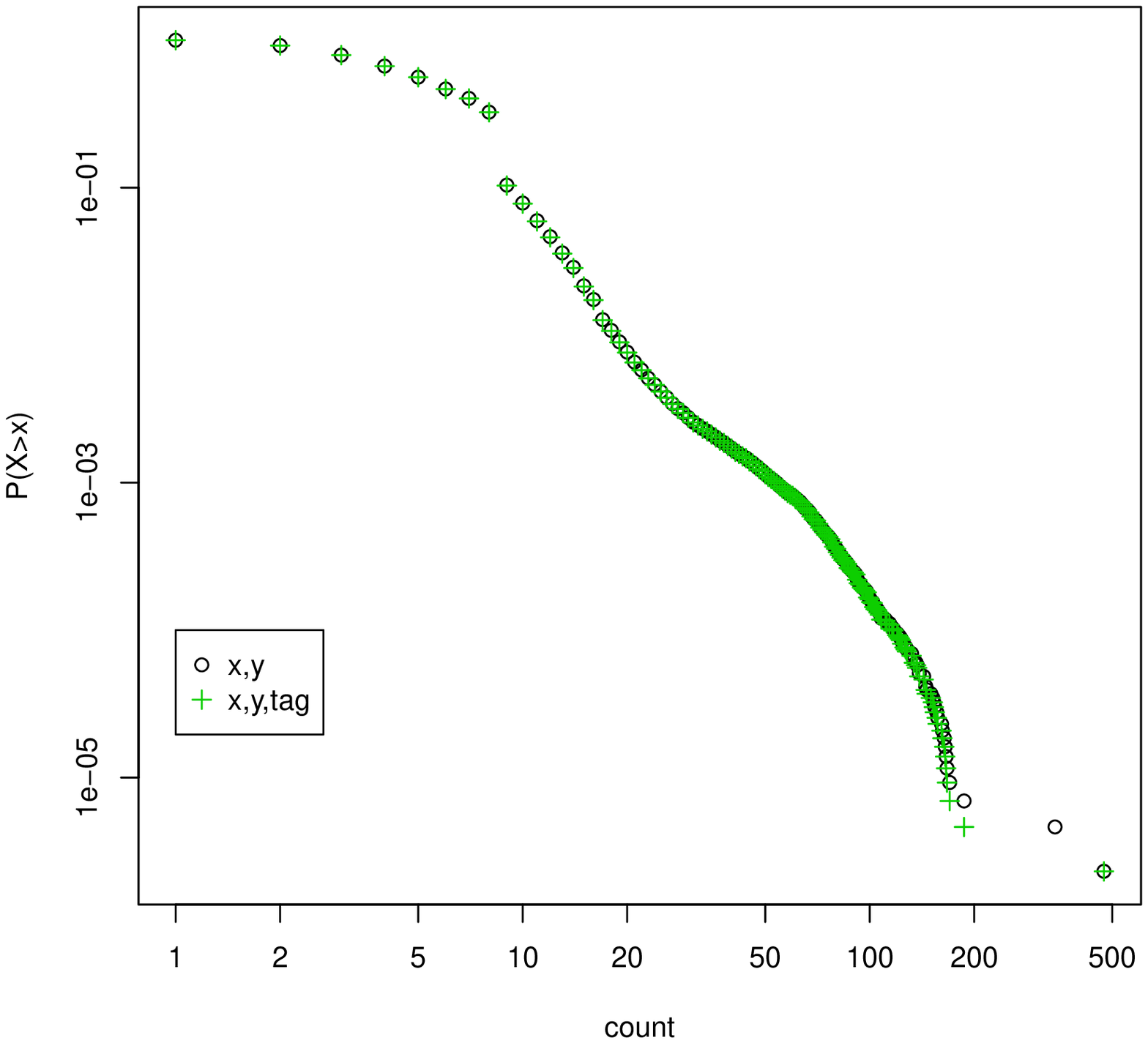}&\includegraphics[scale=.4]{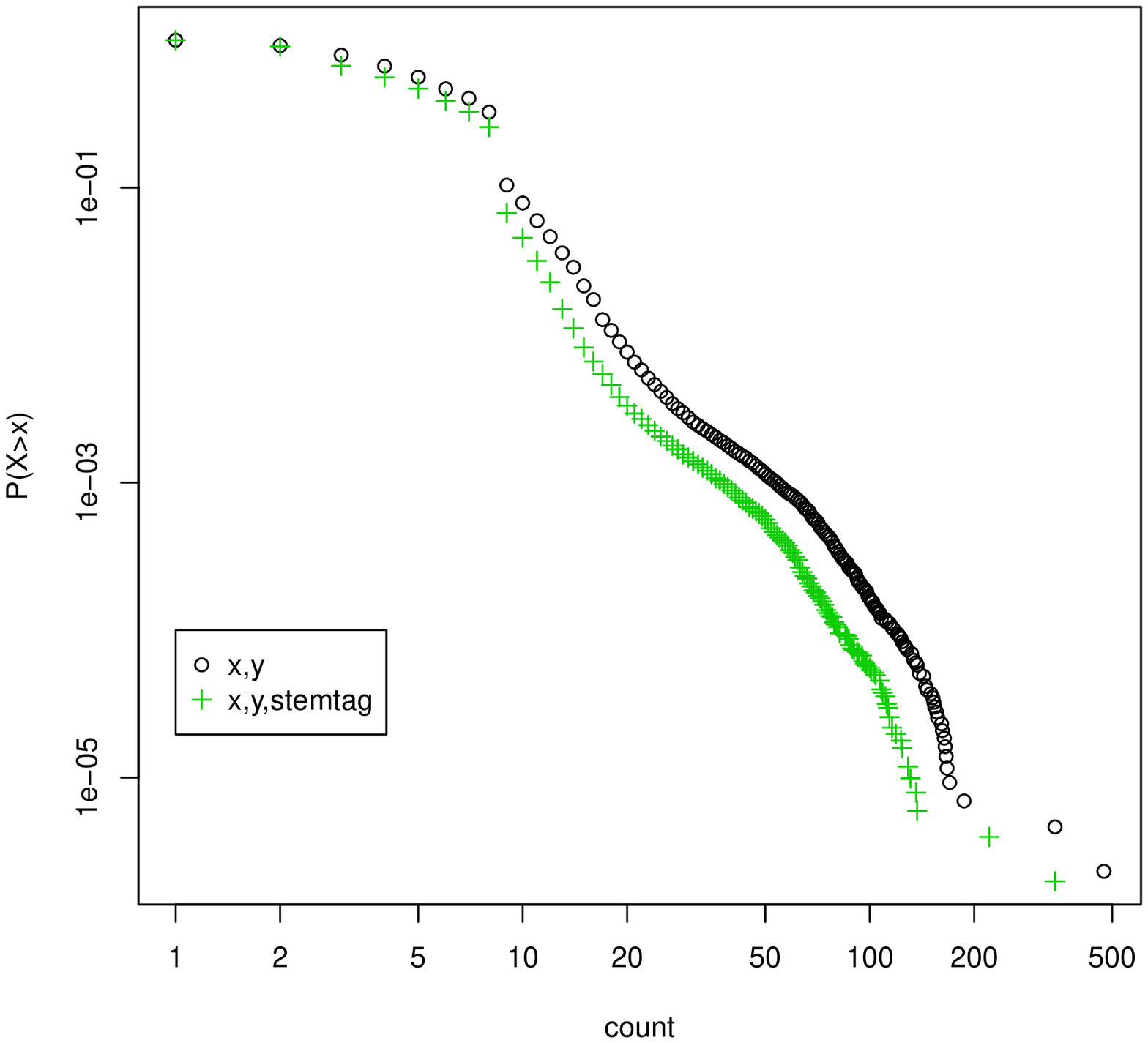}
\end{tabular}
\caption{Tail distributions for $x,y$ and a $3$-element subset of attributes containing $x,y$.}\label{taild}
\end{figure*}

\smallskip
Table \ref{rho} gives the values of $\rho(S,s)$ for projections of the data set onto different attribute sets: first $x,y$, then $x,y$ and a third attribute. 
The extra attribute leading to the closest value of $\rho$ for the attribute set $x,y$ is $\latin$. 
Figure \ref{taild} shows the tail distributions 
of $x,y$ and $x,y$ plus an additional attribute. It can be seen that adding $\latin$ to $x,y$ does not change the tail distribution, as could be expected from the $\rho$ values.
All other such attribute sets, with the exception of $x,y,\Tag$, yield tail distributions which are different from that of $x,y$.
This means that $\rho(S,s)$ is able to distinguish between point clouds with different tail distributions, and the example of $x,y,\latin$ and
$x,y,\Tag$ shows that $\rho(S,s)$ can distinguish more point clouds than the tail distribution.
A double logarithmic plot of the tail distribution reveals whether a uniform distribution is likely or not, 
because the uniform distribution yields a plot displaying a longer `survival time' at probability $1$, followed by a sharp drop to $0$.
As they are almost straight lines, a heavy-tailed distribution like a power law or a log-normal distribution might be likely.

\begin{table}
\centering
\begin{tabular}{|c|c|c|}\hline
log-normal&\multicolumn{2}{c|}{log-normal vs.\ power law}\rule{0pt}{3.5mm}\\
$p$-value&log-likelihood ratio&$p$-value\rule{0pt}{3.5mm}\\\hline
0.77&0.84&0.20\rule{0pt}{3.5mm}\\\hline
\end{tabular}
\caption{Plausibility of assuming a log-normal distribution for $x,y$, and comparison with the power law distribution.}\label{lnormtest}
\end{table}

\smallskip
The framework in \cite{CSN2009} shows how  to perform a goodness-of-fit test of heavy-tailed distributions  together with a
 method for directly comparing two alternative distributions. In \cite{heavyTails} a brief overview of these methods is given. 
Table \ref{lnormtest} shows the results of a test for a log-normal distribution for the $x,y$ attribute set and a comparison with the power law distribution. As the $p$-value for the first test is high, a log-normal distribution cannot be ruled out. The log-likelihood ratio being positive shows that the log-normal distribution is preferred over the power law, and as the corresponding $p$-value is low this time, statistical fluctuations are not very likely to be the cause of this test result. This strengthens the hypothesis that the underlying distribution may be log-normal.

\section{Conclusions}

The distribution of a point cloud has a strong influence on the structure of an optimal tree-index for accessing it. When using Hilbert curves or their higher-dimensional generalisations which are called Gray-Hilbert curves (as they use the Gray code), then the optimal tree-index is given by the smallest sub-tree of the Gray-Hilbert tree whose leaves are the data points. It can be seen from the results that this \emph{scaled} Gray-Hilbert curve index has the potential of saving a lot of superfluous  nodes which are inevitable in the static Gray-Hilbert curve index, especially in higher dimension. 
This means that a large amount of the space-filling curve keys can be economised with this index.
The Gray-Hilbert local sparsity measure is able to distinguish point clouds with differing tail distributions, and moreover can distinguish more than the tail distribution function. It would be interesting to find out how the distinguishing properties of data sets can be expressed by other means.

\section*{Acknowledgements}
The authors of this work 
intend to acknowledge something in the published version of this article. They would also like to express their thanks to 
 the Center for Tropical Forest Science of the Smithsonian Tropical Research Institute
for providing the data set.

{
  \bibliographystyle{plain}
		\normalsize
		\bibliography{biblio} 
}

\end{document}